\documentclass[fleqn,usenatbib]{mnras}

\usepackage{newtxtext,newtxmath}

\usepackage[T1]{fontenc}
\usepackage{ae,aecompl}

\usepackage{graphicx}	
\usepackage{amsmath}	
\usepackage{amssymb}	
\usepackage{color}
\usepackage{bm}
\hypersetup{colorlinks,
	citecolor=[rgb]{0.1 , 0.3, 0.8}}

\newcommand{\lya}{\mbox{${\rm Ly}\alpha$}}

\newcommand{\Avcrit}{\ensuremath{A(V)_{\rm crit}}}

\newcommand{\zabs}{\ensuremath{z_{\rm abs}}}
\newcommand{\sigdla}{\ensuremath{\sigma_{\textsc{dla}}}}

\newcommand{\kz}{\ensuremath{\kappa_{\textsc z}}}

\newcommand{\HI}{\ion{H}{i}}
\newcommand{\NHI}{\ensuremath{N_{\rm H\, \textsc{i}}}}
\newcommand{\NHIcen}{\ensuremath{N_{\rm H\, \textsc{i},\, 0}}}
\newcommand{\OmegaDLA}{\ensuremath{\Omega_{\textsc{dla}}}}

\newcommand{\logNHI}{\ensuremath{\log({\rm N_{H\, \textsc{i}}})}}
\newcommand{\logNHIcm}{\ensuremath{\log({\rm N_{H\, \textsc{i}}\: /\: cm}^{-2})}}
\newcommand{\sigmaHI}{\ensuremath{\sigma_{\textsc{hi}}}}

\newcommand{\result}[3]{\ensuremath{#1^{+#2}_{-#3}}}
\newcommand{\q}[1]{`#1'}

\title[High-$z$ DLA Galaxy Model]{High-redshift Damped Ly$\alpha$ Absorbing Galaxy Model Reproducing the $\NHI-Z$ Distribution}

\author[J.-K. Krogager et al.]{
	Jens-Kristian Krogager$^{1}$,
	Palle M{\o}ller$^{2}$,
	Lise B. Christensen$^{3}$,
	Pasquier Noterdaeme$^{1}$,\newauthor
	Johan P. U. Fynbo$^{3}$,
	Wolfram Freudling$^{2}$\\
$^{1}$Institut d'Astrophysique de Paris, CNRS-SU, UMR7095, 98bis bd Arago, FR-75014 Paris, France; {E-mail: krogager@iap.fr} \\
$^{2}$European Southern Observatory, Karl-Schwarzschildstrasse 2, D-85748 Garching bei M{\"u}nchen, Germany\\
$^{3}$DARK, Niels Bohr Institute, University of Copenhagen, Lyngbyvej 2, DK-2100 Copenhagen {\O}, Denmark
}

\begin{document}

\label{firstpage}
\pagerange{\pageref{firstpage}--\pageref{lastpage}}
\maketitle

\begin{abstract}
We investigate how damped Lyman-$\alpha$ absorbers (DLAs) at $z \sim 2-3$, detected in large optical spectroscopic surveys of quasars, trace the population of star-forming galaxies. Building on previous results, we construct a model based on observed and physically motivated scaling relations in order to reproduce the bivariate distributions of metallicity, $Z$, and \ion{H}{i} column density, \NHI. Furthermore, the observed impact parameters for galaxies associated to DLAs are in agreement with the model predictions. The model strongly favours a metallicity gradient, which scales with the luminosity of the host galaxy, with a value of $\gamma^* = -0.019 \pm 0.008$~dex~kpc$^{-1}$ for $L^*$ galaxies that gets steeper for fainter galaxies. We find that DLAs trace galaxies over a wide range of galaxy luminosities, however, the bulk of the DLA cross-section arises in galaxies with $L \sim 0.1~L^*$ at $z\sim 2.5$ consistent with numerical simulations.
\end{abstract}

\begin{keywords}
\vspace{-2mm}
	galaxies: high-redshift
	--- galaxies: statistics
	--- quasars: absorption lines
\end{keywords}


\section{Introduction}
\vspace{-2mm}
	The properties of neutral gas are crucial for a complete understanding of galaxy evolution. Locally, the neutral gas can be studied directly through the \HI\ 21-cm transition \citep[e.g.,][]{Zwaan2005, Walter2008}. At high redshift the neutral gas phase is best studied through \lya\ absorption towards background quasars. The so-called damped \lya\ absorbers (DLAs) with $\NHI > 2\times 10^{20}$~cm$^{-2}$ are of particular interest for studies of galactic environments \citep{Wolfe2005}. Due to their high column density of neutral hydrogen, DLAs arise in predominantly neutral gas \citep{Viegas1995}. As a result, ionization corrections for metallicity measurements are negligible.
	
	The distribution function of \NHI\ for DLAs, $f(\NHI)$, has been studied in detail \citep{Prochaska2009, Noterdaeme2012c, Bird2017} and provides a measurement of the cosmic mass density of neutral hydrogen in DLAs, \OmegaDLA. It is found that \OmegaDLA\ makes up $\sim$80\,\% of the total mass density of \HI\ at $z>2.2$ \citep{Noterdaeme2009b}.
	
	The plethora of low-ionization metal absorption lines make it possible to obtain accurate measurements of the metallicity, $Z$, in the neutral gas \citep{Rafelski2012, Jorgenson2013, DeCia2018}. It is found that the average $Z$ increases towards lower redshifts in agreement with expectations from the build up of metals via star formation \citep{Dvorkin2015}. 
	
	In order to properly interpret the properties of DLAs it is important to know which galaxies give rise to DLAs over cosmic time.
	One way to understand the galaxy population associated to DLAs is through numerical simulations. Large cosmological simulations are able to match the observables from DLAs at $z\approx3$ fairly well \citep[e.g.][]{Pontzen2008, Altay2013, Bird2013, Bird2014, Bird2015, Rahmati2014}. However, the redshift evolution and the details of the bivariate distribution of \NHI\ and $Z$ are still not well-understood \citep{Hassan2020} and the properties of DLAs in simulations depend strongly on the rather ad-hoc feedback mechanisms from supernovae and quasars assumed in the simulations.
	
	Another way to study DLA galaxy properties is by cross-correlation analyses with \lya\ forest absorbers in order to measure the DLA bias, $b_{\textsc{dla}}$ \citep{FontRibera2012, Perez-Rafols2018a}. Based on a large statistical analysis, \citet{Perez-Rafols2018a} find an average bias of $b_{\textsc{dla}}=2.0\pm0.1$, which translates to a rather large halo mass ($\sim 10^{11}$~M$_{\odot}$) if all DLAs reside in halos of the same mass; However, the inferred DLA halo mass depends on the distribution function for DLA cross-section as a function of halo mass. A power-law scaling between DLA cross-section and halo mass with index larger than unity implies that DLAs instead reside in a large range of halo masses, where the minimum mass depends critically on the assumed power-law index \citep{Perez-Rafols2018a}. Moreover, the inferred $b_{\textsc{dla}}$ depends on metal line strength \citep{Perez-Rafols2018b} indicating that more metal-enriched DLAs reside in more massive halos, consistent with the mass--metallicity relation inferred for DLAs \citep{Moller2013, Christensen2014}.
	
	The most direct method to examine the galaxies associated to DLAs is through direct detections \citep[see compilation by][]{Moller2020}. Yet, detections of high-$z$ DLA-galaxies have been scarce due to their intrinsically faint nature \citep{Fynbo1999, Haehnelt2000, Schaye2001, Krogager2017}. Recently, however, powerful integral field spectrographs enable more effective follow-up of DLA-galaxies by probing large areas around the background quasar \citep{Fumagalli2017}.
	
	Since we only observe the brightest DLA-galaxies, it is necessary to extrapolate from the individual associations to obtain the global properties of DLAs. \citet{Fynbo2008} have carried out a successful modelling approach to reproduce the $Z$ and impact parameter distribution of DLAs at $z\approx3$. \citet{Padmanabhan2017} have studied the \HI\ distribution of DLAs using an analytical formalism to link halo properties to \HI\ mass and cross-section. Their model reproduces well the redshift evolution of ${\rm d}n_{\textsc{dla}}/{\rm d}z$ but overproduces the number of high \NHI\ systems.
	
	While the models by \citet{Fynbo2008} and \citet{Padmanabhan2017} are successful at predicting the distributions of $Z$ and \NHI\ independently, so far there have been no attempts to model both of these key properties simultaneously.
	Here we therefore extend the original model by \citet{Fynbo2008} to include a statistical prescription for \NHI\ in order to describe the bivariate \NHI-$Z$ distribution. We furthermore include the effects of a dust bias in optical quasar selection affecting the observed DLA properties \citep{Pei1991, Murphy2016, Krogager2019}. In this paper, we perform a Bayesian analysis to constrain the model parameters including priors on parameters that have already been constrained independently.

	The paper is organized as follows: In Sect.~\ref{data}, we describe the compilation of data used to constrain the model; in Sect.~\ref{model},
	we present the details of our model and the parameter estimation; We discuss the results and implications of our work in Sect.~\ref{discussion}; and lastly, we summarize our findings in Sect.~\ref{summary}.
	
	Throughout this paper, we assume a flat $\Lambda$CDM cosmology with
	$H_0=68\, \mathrm{km s}^{-1}\mathrm{Mpc}^{-1}$, $\Omega_{\Lambda}=0.69$
	and $\Omega_{\textsc{m}} = 0.31$ \citep{Planck2016}.

\vspace{-5mm}
\section{Literature Data}
\label{data}

	In order to constrain our model, we use the observed \NHI\ distribution function derived by \citet{Noterdaeme2012c} based on a statistical sample of $\sim$3500 DLAs at redshifts $2 < z < 3$ detected in $\sim$40000 quasar spectra. However, only a small subset of these have robust measurements of metallicity, $Z$. We compile a sample of $Z$ measurements from the literature combining the three largest samples available by \citet{Quiret2016}, \citet{Jorgenson2013} and \citet{Rafelski2012}. We remove measurements based on limits and those based only on iron as this element tends to deplete heavily onto dust grains. We furthermore only consider DLAs in the redshift range $2 < z < 3$. This sample of 178 DLAs will hereafter be referred to as the \q{full DLA sample}. The selection effects are discussed in more detail in Sect.~\ref{discussion}.
	All values of metallicities are in units of Solar metallicity, $Z_{\odot}$, unless stated otherwise. We have corrected all the measurements described above to the same Solar reference values using measurements by \citet{Asplund2009} and the recommendations by \citet{DeCia2016} as to whether photospheric or meteoritic values are used.

	In order to compare impact parameter predictions from our model, we use the sample of high-redshift ($\zabs \gtrsim 2$) DLAs with confirmed emission counterparts compiled by \citet{Moller2020} and \citet{Krogager2017}. We restrict the sample to the subset with $\logNHIcm > 20.3$. To this sample, we add three counterparts reported by \citet{Ranjan2020} at $z \sim 2.3$ together with one DLA counterpart by \citet{Srianand2016} at $z=3.247$ and one by \citet{Fumagalli2017} at $z=3.25$. This sample will hereafter be referred to as the \q{DLA galaxy sample}.

\section[Modelling High-z DLAs]{Modelling High-Redshift DLAs}
\label{model}

	The model described here is based on the work by \citet{Fynbo2008}. We here offer a short summary of the model framework and refer the reader to the original work for further details. In this work, we only consider redshifts between $2 < z < 3$. The model uses a selection probability of DLAs given by $P_{\textsc{dla}} \propto \sigdla \ \phi(L)$, where \sigdla\ is the effective cross-section of DLAs and $\phi$ is the UV luminosity function. For the luminosity function, a Schechter function of the form $\phi(L) = \phi_0\, (L/L^*)^{\alpha} \, \exp(-L/L^*)$ is used.

	We assume \sigdla, given by $\pi R_{\textsc{dla}}^2$, to scale with luminosity, through a Holmberg relation $R_{\textsc{dla}} = R^*_{\textsc{dla}} (L/L^*)^t$, and vanish below a limiting luminosity $L_{\rm min}$. In what follows, all quantities marked by $^*$ are referring to the given quantity of an $L^*$ galaxy, e.g., the radial extent of DLA cross-section for an $L^*$ galaxy is denoted $R^*_{\textsc{dla}}$.
	The absolute value of $R^*_{\textsc{dla}}$ is obtained by requiring that the incidence rate, ${\rm d}n_{\textsc{dla}}/{\rm d}z$, matches the observed value at $z \approx 2.5$. We calculate ${\rm d}n/{\rm d}z$ as:
	\begin{equation}
		\frac{{\rm d}n}{{\rm d}z} = \pi {R^*_{\textsc{dla}}}^2\, \phi_0\, c\, (1+z)^2\, H^{-1}(z) \int_{L_{\rm min}}^{\infty} L^{\alpha + 2t}\, e^{-L}\, {\rm d} L~,
	\end{equation}
	
	\noindent
	where $L$ is in units of $L^*$ and the Hubble parameter is given as:
	\begin{equation}
		H(z) = H_0 \sqrt{\Omega_{\textsc{m}} (1+z)^3 + \Omega_{\Lambda}}~.
	\end{equation}
	
	\noindent
	We here use the observed value of ${\rm d}n_{\textsc{dla}}/{\rm d}z = 0.21\pm0.04$ at $z=2.5$ \citep{Zafar2013b}.

	Galaxies are then sampled from the luminosity function weighted by \sigdla, and an impact parameter, $b$, is drawn randomly with a probability $P(b) \propto b$ for $b\leq R_{\textsc{dla}}(L)$, that is, the probability is weighted by area.

	A central metallicity, $Z_0$ is assigned to each galaxy assuming a metallicity--luminosity ($Z-L$) relation: $\log(Z_0) = \log Z_0^* + \beta \times M_{\textsc{uv}}$.
	A radial metallicity gradient is then assumed in order to obtain a value of the metallicity, $Z_{\rm abs}$, at the impact parameter where the absorption system would be observed. This gradient is taken to be luminosity dependent with a variable power-law index:
	$\gamma = \gamma^*\, L^{q_{\rm z}}$.
	While the original work by \citet{Fynbo2008} assumed a fixed value of $q_{\rm z} = -t$ following \citet{Boissier2001}, we keep this index as a free parameter in order to quantify whether a universal gradient (i.e., $q_{\rm z} = 0$) is preferred over a luminosity dependent gradient.

\subsection{Including \HI\ and dust}
\label{model:NHI}
	The neutral gas in galaxies is roughly expected to follow an exponential distribution with radius: $\NHI(r) = \NHIcen\, \exp(-r/r_{\textsc{hi}})$ \citep{Walter2008}. The scale length, $r_{\textsc{hi}}$, is calculated by demanding that $\NHI(r=R_{\textsc{dla}}) = 2\times 10^{20}$~cm$^{-2}$. For this reason, $r_{\textsc{hi}}$ is not a free parameter in this model. The central \NHI\ value is however kept as a free variable with an adopted fiducial value of $\NHIcen=10^{22}$~cm$^{-2}$.

	Motivated by observations of local \HI\ discs by the THINGS survey \citep{Walter2008}, we include a stochastic term in the radial \NHI\ distribution to account for local fluctuations in column density as well as inclination effects which are not explicitly modelled. The fluctuations are implemented as a log-normal scatter around the smooth average radial profile:
	$$ \log\NHI(r) = \log\NHIcen - \frac{\log(e)}{r_{\textsc{hi}}} r + \mathcal{N}(0, \sigmaHI)~,$$

	\noindent
	where the log-normal scatter \sigmaHI\ is a free variable in our model with a fiducial value of 0.3~dex.
	
	Since we now have a prescription for both \NHI\ and $Z$, we can calculate the expected amount of optical dust extinction along the absorption sightline, $A(V)$. This value is obtained following \citet{Zafar2019} assuming a constant dust-to-metals ratio, $\log\kz = -21.4$.
	
	Lastly, we include a dust bias to account for the fact that quasars behind dusty DLAs are systematically under-represented due to the complex colour and magnitude selection criteria. We calculate the selection probability as $P_{\textsc{qso}} = {\rm sech}(x^2)$, where $x=A(V)/\Avcrit$. This functional form fits very well the calculated selection probability by \citet{Krogager2019} for a value of $\Avcrit=0.25$~mag. The model distributions are filtered according to $P_{\textsc{qso}}$ to produce a mock observable model distribution.
	
	In total, the original model contains 9 free parameters:
	$\{\phi_0\,,\ M_{\textsc{uv}}^*,\ \alpha,\ L_{\rm min}\,,\ t,\ \beta,\ \gamma^*,\ Z^*,\ q_{\rm z} \}$,
	and with the above modifications, we have effectively added the following 3 parameters: $\{\NHIcen\,,\ \sigmaHI\,,\ \Avcrit\, \}$.

\subsection{Constraining model parameters}

	Due to the significant degeneracies in the parameters we constrain the model parameters using a Bayesian approach. This also allows us to include priors since we have independent constraints on many parameters. For this purpose, we use the Python package {\sc Emcee} \citep{Foreman-Mackey2013}.
	
	We use the observed distributions of \NHI\ and $Z$ (see Sect.~\ref{data}) to statistically constrain the model parameters. For a given set of parameters, we draw a large sample of 50,000 DLAs and then calculate the model $f(\NHI)$ in the same bins as the data, using only DLAs that pass the mock quasar selection as implemented here using $P_{\textsc{qso}}$ (see Sect.~\ref{model:NHI}). The distribution function is normalized by requiring that the integral
	$\int_{N_{\textsc{dla}}}^{\infty} f(\NHI) {\rm d}N$
	matches the observed value of $\int_{N_{\textsc{dla}}}^{\infty}f(\NHI)\,{\rm d}N$, where $N_{\textsc{dla}}=2\times10^{20}$~cm$^{-2}$. This normalization ensures a correct absolute scaling of $f(\NHI)$ in order to match the observed ${\rm d}n_{\textsc{dla}}/{\rm d}z$. We then calculate the likelihood assuming Gaussian statistics given the uncertainties quoted by \citet{Noterdaeme2012c}.
	
	Similarly, we obtain a model distribution for $Z$ which we compare to the observed distribution using a Kolmogorov--Smirnov (KS) test. Although the $p$-value from a KS test is not an exact estimator of the formal likelihood, the two are sufficiently correlated \citep{Krueger2017} allowing us to use $P_{\textsc{ks}}$ as an estimate of the likelihood. In our case we find that the $p$-value provides tighter constraints than other likelihood estimators (such as kernel density estimators).
	
	Along with the constraints from the \NHI\ and $Z$ distributions, we have the following independent constraints:
	
	\vspace{-1mm}
	\begin{enumerate}
	
		\item The average reddening for DLAs, which pass the optical quasar selection criteria, $\langle E(B-V)\rangle_{\rm obs}$, must not exceed 21~mmag at the 3-$\sigma$ level \citep{Murphy2004}. We use this conservative upper limit since there is significant disagreement among various measurements \citep[see][]{Murphy2016};
		\vspace{1mm}
		
		\item The average metallicity of extremely strong DLAs (ESDLAs, $\logNHI > 21.7$, \citealt{Noterdaeme2014}) is observed to be $\langle Z \rangle = -1.30 \pm 0.05$ \citep{Ranjan2020}.

	\end{enumerate}
	
	\vspace{-1mm}
	The joint likelihood is then taken as the product of the independent likelihoods taking into account the higher number of degrees of freedom for the \NHI\ data.

	\subsubsection{Priors}
	All priors used in our statistical analysis are summarized in Table~\ref{tab:pars}. 
	For parameters where we have no prior knowledge we use flat priors over a reasonable range of parameter space. We have verified that the choice of prior ranges do not affect the results and all values are constrained well within the chosen ranges. For the parameters with more restrictive priors, the details of the priors are given below.

	We constrain the shape of the luminosity function following observations from \citet{Malkan2017}. The parameters $M_{\textsc{uv}}^* = -20.9$ and $\phi_0 = 1.7\times 10^{-3}$~Mpc$^{-3}$ are kept fixed as they agree very well from one survey to another \citep[table 3]{Malkan2017}. On the other hand, we choose to keep $\alpha$ as a free parameter since this value shows large dispersion among various surveys. The average value and the standard deviation are used as a prior on $\alpha$.
	In this work, we adopt a fiducial value of $L_{\rm min} = 10^{-4}L^*$. This value is consistent with numerical simulations \citep{Bird2013} and semi-analytical modelling \citep{Dvorkin2015}.
	
	The slope of the $Z-L$ relation, $\beta$, is constrained from observations of galaxies at $z \sim 0.5-1$ \citep{Kobulnicky2004, Hidalgo2017}. We obtain an average value of $\langle\beta\rangle = 0.21$\footnote{Note the change of sign in our definition with respect to \citet{Hidalgo2017}.} (weighted by individual uncertainties) with a standard deviation of $0.05$. This average value is taken as our prior and is in agreement with results from previous modelling \citep{Krogager2017}. Although the redshift range studied by \citet{Hidalgo2017} is lower than what we try to model here, there is evidence that the slope of the related mass--metallicity relation does not evolve significantly with redshift \citep{Maiolino2008}. It is therefore reasonable to assume that the slope of the metallicity--luminosity relation would also remain constant with redshift.

	The normalization of the $Z-L$ relation, $Z^*$, is constrained from observations of the mass--metallicity relation at $z \sim 2.2$ \citep{Maiolino2008}. We find that galaxies around $M^*$ have roughly Solar metallicity. In order to take into account the observed scatter as well as systematics, we use a weak prior on $Z^*$: $\log Z^* = 0.0 \pm 0.2$.
	
	We use a fiducial value of $A(V)_{\rm crit} = 0.25$~mag, derived using the calculation of selection probability as a function of $A(V)$ from \citet{Krogager2019}. The value of 0.25~mag corresponds to a limiting $A(V)$ of 0.57~mag for a selection probability of $P_{\rm QSO} = 0.01$, i.e., DLAs with $A(V)$ larger than 0.57~mag have a selection probability less than 1\% in SDSS-II (up until DR7). Since the \q{full DLA sample} is observed with larger telescopes than the SDSS this $A(V)$ limit may differ with respect to the fiducial value. We therefore keep this value as a free parameter and assign a rather arbitrary logarithmic uncertainty for the prior of 0.3~dex. The results do not depend strongly on the chosen width of the prior distribution.\\

	Using the priors mentioned above, we obtain an initial estimate of the parameters using {\sc Emcee} to explore parameter space with 100 walkers for 600 steps. The posterior probability distribution shows a strong one-to-one anti-correlation between the parameters $t$ and $q_{\rm z}$ with a Spearman correlation coefficient of $-0.92$.
	A similar anti-correlation is implemented in the original model by \citet{Fynbo2008} following \citet{Boissier2001}, who use $q_{\rm z} = -t$.
	Based on the observed anti-correlation we adopt the constraint: $q_{\rm z} = -t$. Hence, $q_{\rm z}$ is no longer considered a free parameter.
	
	Although the average metallicity gradient of DLA hosts has been inferred by \citet{Christensen2014}, we do not include this constraint as a prior on $\gamma^*$. Since these authors have assumed a constant gradient with no luminosity dependence and no separation between low- and high redshift DLA galaxies, including their obtained metallicity gradient as a prior in this work could possibly bias our results.

	\renewcommand{\arraystretch}{1.5}
	\begin{table}
		\begin{center}
		\caption{Summary of free model parameters}
		\label{tab:pars}
		\begin{tabular}{lcrr}
			\hline
			Parameter  &  Prior $^a$  &  Best-fit $^b$  &  Ref.  \\
			\hline
			$\alpha$                    &  $-1.70 \pm 0.20$    &  \result{-1.70}{0.20}{0.20}     & (1)  \\
			$t = -q_{\rm z}$            &  [\,$-5$ , $+5$\,]   &  \result{0.51}{0.27}{0.17}      &      \\  
			$\beta$                     &  $0.21 \pm 0.05$     &  \result{0.20}{0.04}{0.03}      & (2)  \\
			$\gamma^*$ / dex~kpc$^{-1}$ &  [\,$-1$ , $+1$\,]   &  \result{-0.019}{0.008}{0.008}  &     \\
			$\log Z_0^* / Z_{\odot}$    &  $0.0 \pm 0.2$       &  \result{-0.02}{0.18}{0.18}      & (3)  \\
			$\log(\NHIcen$ / cm$^{-2})$ &  [\,20 , 24\,]       &  \result{20.91}{0.29}{0.27}     &   \\
			$\sigmaHI$                  &  [\,0 , 2\,]         &  \result{0.54}{0.06}{0.09}      &   \\
			$\log\Avcrit$               &  $-0.6 \pm 0.3$      &  \result{-0.55}{0.22}{0.20}     & (4) \\
			\hline
			Fixed parameters:    &  &  Value  &  Ref. \\
			\hline
			$\phi_0$ / Mpc$^{-3}$       &  &   $1.7\times 10^{-3}$  &  (5) \\
			$M_{\textsc{uv}}^*$         &  &   $-20.9$    &  (5) \\
			$L_{\rm min}$ / $L^*$       &  &   $10^{-4}$  &    \\
			$\log\kz$                   &  &   $-21.4$    &  (6)  \\
			\hline
		\end{tabular}
		\end{center}
		{\flushleft 
		$^a$ Gaussian priors are given as $\mu \pm \sigma$; Flat priors are given as [min , max].\\
		$^b$ Best-fit values are stated as the median value with 16-th and 84-th percentiles as confidence intervals.\\
		
		{\bf References for priors:}
		(1) \citet{Malkan2017};
		(2) \citet{Hidalgo2017};
		(3) \citet{Maiolino2008};
		(4) \citet{Krogager2019};
		(5) \citet{Malkan2017};
		(6) \citet{Zafar2019}.
		}
	\end{table}

	\subsection{Results}
	We obtain the best-fit solution using {\sc Emcee} with 100 walkers for 800 iterations of which we discard the first 200 iterations for which the ensemble has not converged.
	The values of the optimized model parameters are given in Table~\ref{tab:pars}. These are stated as the median value of the posterior probability distribution together with the 16th and 84th percentiles as 1-$\sigma$ confidence intervals.
	
\begin{figure*}
	\includegraphics[width=0.8\textwidth]{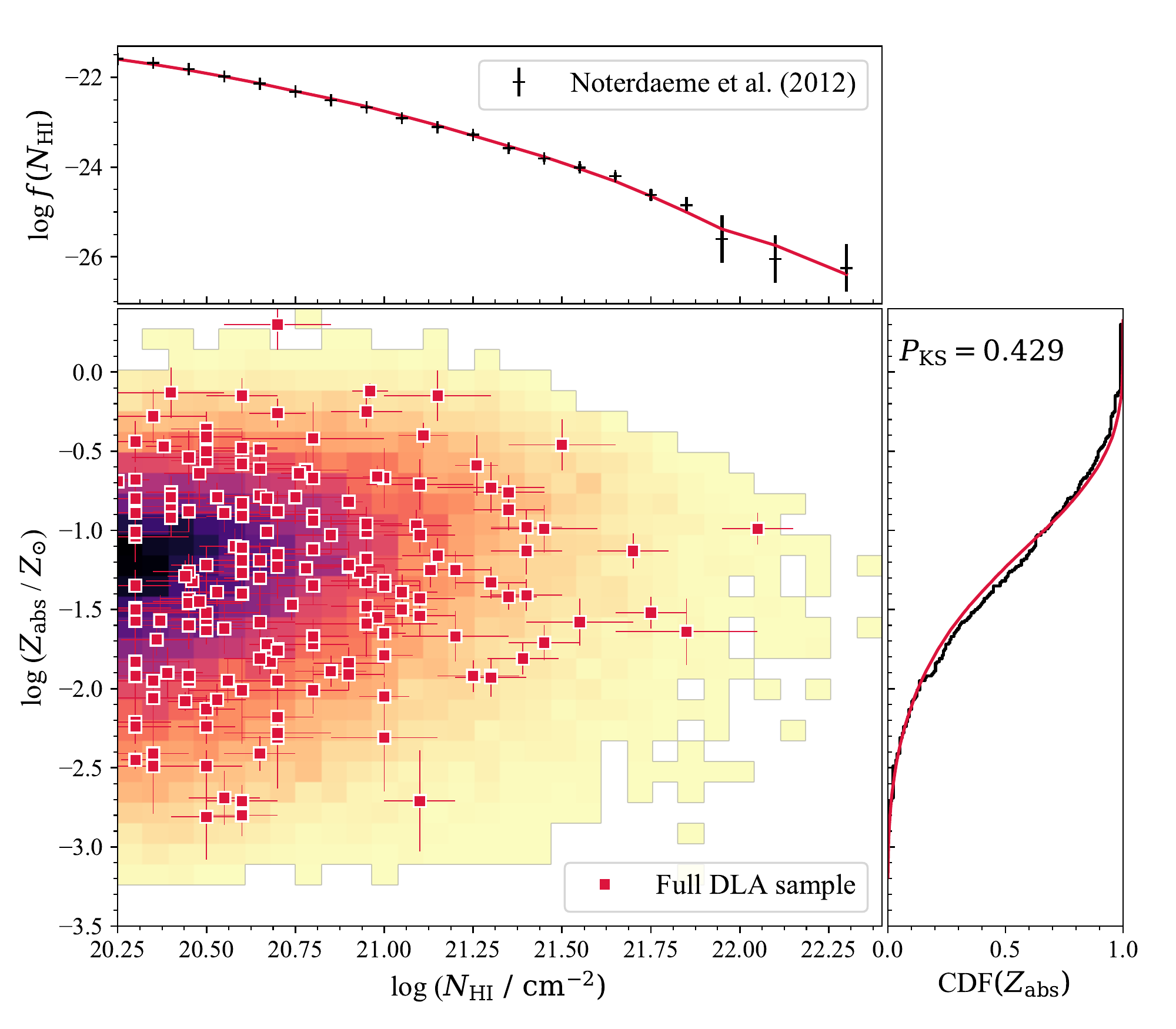}
	\caption{Model prediction for the bivariate \NHI\ and $Z_{\rm abs}$ distribution. $Z_{\rm abs}$ here refers to the metallicity at the given impact parameter in contrast to the central metallicity, $Z_0$, that would be probed by emission line measures. The colors of the model distribution indicates the number of model points in the given bin normalized to a linear scale from 0 to 1. The marginalized $Z_{\rm abs}$ distribution is shown in the right panel as the cumulative distribution function (CDF, in black) together with the best-fit model (in red). The top panel shows the \NHI\ distribution function (black points) compared to the best-fit model (red line).}
	\label{fig:NHI_Z}
\end{figure*}

\begin{figure*}
	\includegraphics[width=0.87\textwidth]{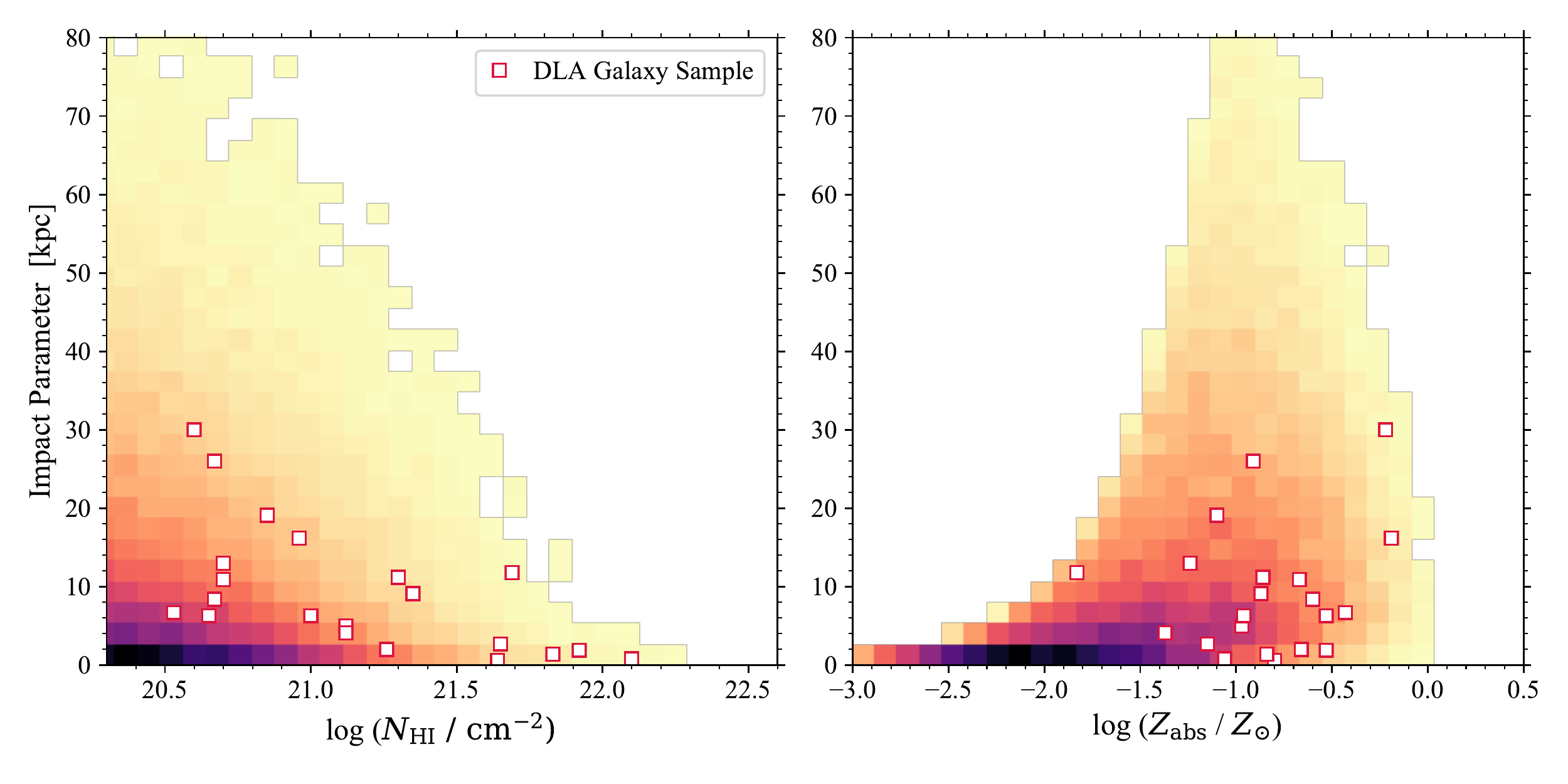}
	\caption{Model prediction for impact parameter as a function of \NHI\ (left) and $Z_{\rm abs}$ (right) for all model galaxies (SFR~$>0$~M$_{\odot}$~yr$^{-1}$). The color scale of the model distributions follows that of Fig.~\ref{fig:NHI_Z}. The sample of DLA galaxies is shown as open, square points.
	We note that the `DLA galaxy' sample is neither complete nor representative and hence should only be compared qualitatively to the underlying model distribution.}
	\label{fig:b_NHI_Z}
\end{figure*}

	The results of the best-fit model are shown in Fig.~\ref{fig:NHI_Z}. We find a very good agreement between the model and the data. For comparison, we also show the impact parameter distribution as a function of \NHI\ and $Z$ in Fig.~\ref{fig:b_NHI_Z}. Since the `DLA galaxy' sample is not complete and suffers from strong and inhomogeneous selection effects (mainly high-metallicity galaxies have been targeted and identified), we do not include the impact parameters in the formal modelling. Nonetheless, it is interesting to compare the model predictions to the observations. We find that the observations indeed overlap with the model predictions which lends qualitative support to the best-fit model, in particular since these observations have not been used to constrain the model parameters.

\section{Discussion}
\label{discussion}

	\subsection{Radial distribution of \NHI}
	The best-fit value of $\NHIcen = 8_{-4}^{+8}\times 10^{20}$~cm$^{-2}$ is consistent with local \HI\ observations from the THINGS survey who report typical values of $\sim10$~M$_{\odot}$~pc$^{-2}$ corresponding to $\sim10^{21}$~cm$^{-2}$ \citep{Walter2008}. The inferred amount of scatter in \logNHI\ is high, $\sigmaHI = 0.54^{+0.07}_{-0.09}$~dex, compared to the fairly smooth radial profiles presented by \citet{Walter2008}.
	This is expected since the individual radial profiles presented by \citet{Walter2008} have been azimuthally averaged. Moreover, the \HI-emission studies provide beam-averaged measurements (typically $100-500$~pc for the THINGS galaxies) which smoothes out small-scale structure. The very small scales probed by quasar sightlines \citep[$\lesssim$1~pc; e.g,][]{Balashev2011} may therefore show much larger local variations.
	A large degree of randomness in \NHI\ is also expected since the neutral medium is highly turbulent \citep{Elmegreen2004} and our model samples the whole galaxy population, not just a single galaxy.

	\subsection{DLA impact parameters}
\begin{figure}
		\includegraphics[width=0.47\textwidth]{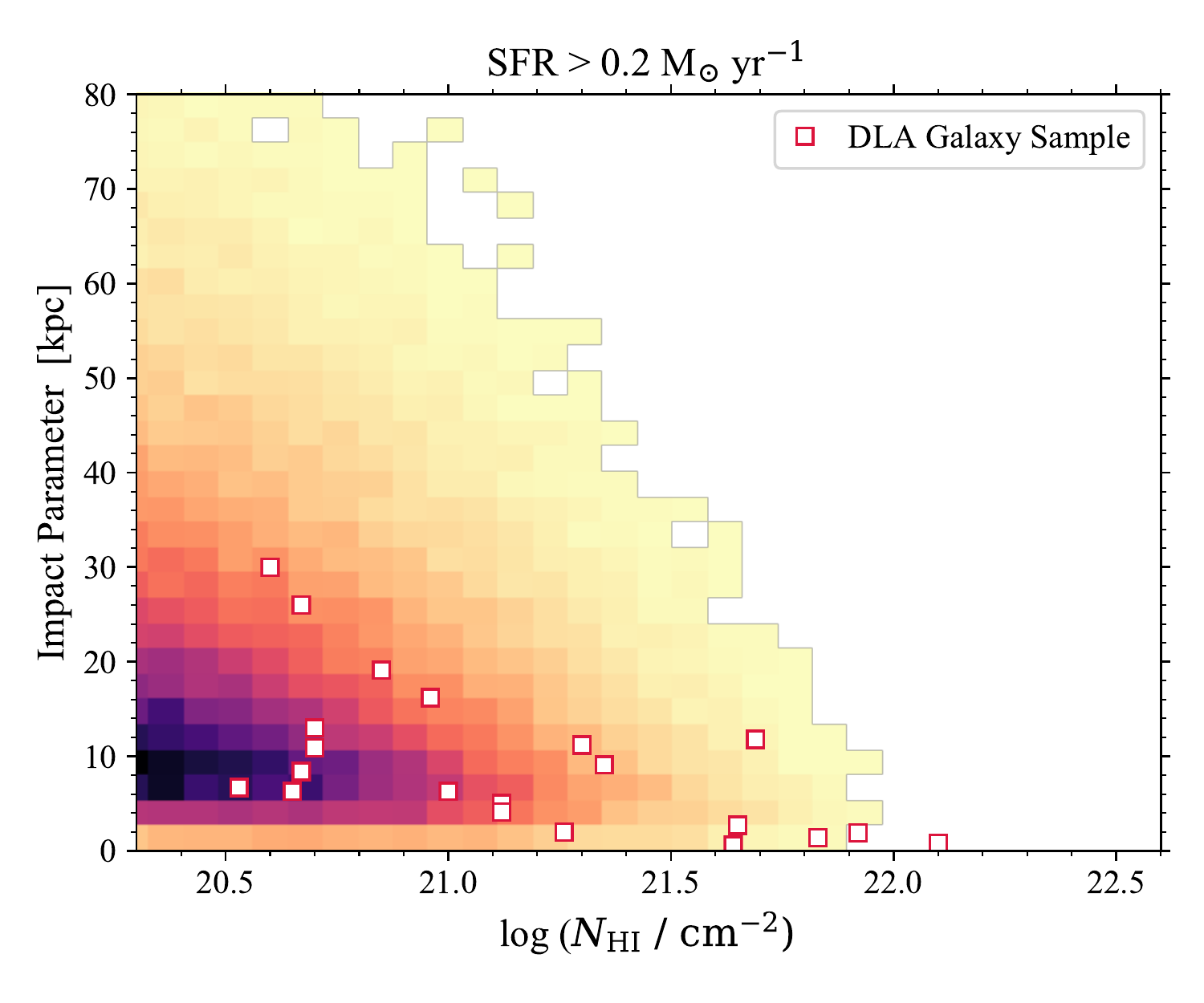}
		\includegraphics[width=0.47\textwidth]{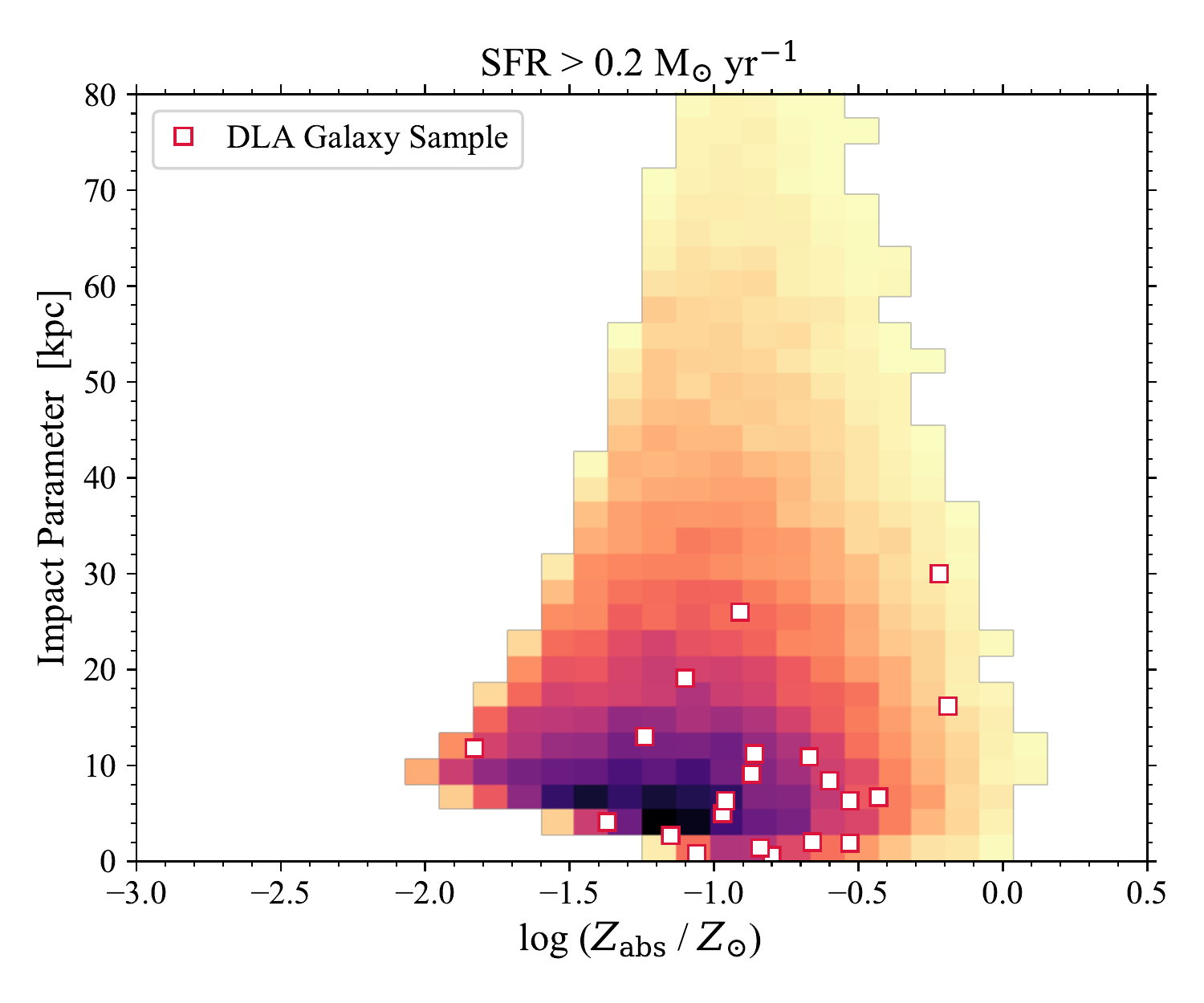}
		\caption{Model prediction for impact parameter as a function of \NHI\ (top) and $Z_{\rm abs}$ (bottom) considering only DLA hosts with ${\rm SFR} > 0.2$~M$_{\odot}$~yr$^{-1}$.}
		\label{fig:bNHI_only}
	\end{figure}

	The distribution of \NHI\ and impact parameter is in qualitative agreement with the simulation by \citet{Rahmati2014}. In order to make a fair comparison, we restrict our model to only consider DLA hosts with similar star formation rates (SFRs) as \citeauthor{Rahmati2014} (${\rm SFR} > 0.004$). For this purpose, we calculate SFR based on the UV luminosity included in our model following \citet{Kennicutt1998}, i.e., ${\rm SFR} \propto L_{\textsc{uv}}$. We find that the median impact parameter increases when looking at DLA hosts with larger SFRs (see top panel of Fig.~\ref{fig:bNHI_only}). This is similar to the simulations by \citet{Rahmati2014}, yet the median impact parameters from our best-fit model are $\sim$2 times larger. The simulations by \citet{Rahmati2014} do not provide explicit information on the absorption metallicity and it is therefore not certain whether they match the bivariate \NHI\--$Z$ distribution.
	
	Our results are in better agreement with the recent simulation by \citet{Rhodin2019}, who find larger impact parameters for high-redshift absorbers. The observed anti-correlation between \logNHI\ and impact parameter is however not recovered at $z>1$ in the simulation by \citet{Rhodin2019}. Their simulation only addresses one Milky Way type progenitor, which is more massive than the average DLA host in our work, and it is therefore difficult to perform a one-to-one comparison between their simulation and our work.

	We have also investigated how the metallicity--impact parameter distribution varies when considering only DLA hosts with SFR larger than 0.2~M$_{\odot}$~yr$^{-1}$. This limit corresponds broadly to the SFR limit obtained in the \q{DLA galaxy sample}.
	In the lower panel of Fig.~\ref{fig:bNHI_only}, we show the distribution of impact parameters as a function of absorber metallicity for DLA hosts with SFR larger than 0.2~M$_{\odot}$~yr$^{-1}$. By restricting the model distribution to this SFR limit we obtain a good match to the observations. However, it is not possible to quantify the agreement in more detail given the inhomogeneous sample selection of the DLA galaxy sample, combined with the fact that the detections are based on different emission lines (e.g., Ly$\alpha$, H$\alpha$, [\ion{O}{iii}]) yielding inhomogeneous detection limits.

	\subsection{DLA cross-section and halo properties}
	In the following, we shall compare the cross-section of DLAs, $\sigma_{\textsc{dla}}$, as a function of their host luminosity to results from numerical simulations. \citet{Bird2014} find that $\sigma_{\textsc{dla}}$ as a function of halo mass is well-reproduced by a power-law with an index between 0.8 and 1. The upper range of their results is in good agreement with the luminosity scaling we infer for $\sigma_{\textsc{dla}}$ of 1.02 ($\sigma_{\textsc{dla}} \propto L^{2t}$, and $t=0.51$) assuming a constant mass-to-light ratio. The normalization of the power-law relation inferred by \citet{Bird2014} is $\sim100$~kpc$^2$ at M$_{\rm h} = 10^{10}$~M$_{\odot}$. 
	\citet{Pontzen2008} find a steeper relation for $\sigma_{\textsc{dla}}$ as function of halo mass for low masses which flattens at masses above $10^{10}$~M$_{\odot}$ and the DLA cross-section at $10^{10}$~M$_{\odot}$ in their simulation is $\sim 50$~kpc$^{2}$.
	As our analysis is based on the DLA galaxy luminosity, we need to assume a halo-mass-to-light ratio in order to compare our model to the simulations. The best-fit model yields a characteristic scale of DLA cross-section for a $L^*$ galaxy of $R_{\textsc{dla}}^* = 31$~kpc. In order to reproduce the simulations by \citet{Pontzen2008} and \citet{Bird2014}, we therefore need to assume ${\rm M_h}/L_{\textsc{uv}} \sim 20-40$~M$_{\odot}/L_{\odot}$ for $L^*$ galaxies.
	This ${\rm M_h}/L_{\textsc{uv}}$ ratio is consistent with what is found in the literature \citep{Vale2006, Mason2015}.
	
	We then compare the distribution of DLA galaxy luminosities from our model to the halo mass distribution from the simulations by \citet{Bird2014}. For our best-fit model, we find that the bulk of the DLA cross-section is contributed by galaxies around $0.1~L^*$. Assuming the average inferred ${\rm M_h}/L_{\textsc{uv}} = 30$, we obtain a bulk halo mass for DLAs of $6\times10^{10}$~M$_{\odot}$. This is consistent with the peak of the halo-mass distribution presented by \citet[their fig. 2]{Bird2014}.
	The minimum halo mass contributing to the DLA cross-section in the simulations by \citet{Bird2014} is $M_{\rm min} \approx 10^9$~M$_{\odot}$ consistent with modelling studies of the halo properties of DLAs \citep[e.g.,][]{Pontzen2008, Barnes2009, Barnes2014, Padmanabhan2017}. Using the ${\rm M_h}$--$L_{\textsc{uv}}$ relation by \citet{Mason2015}, we find that a minimum halo mass of $10^9$~M$_{\odot}$ corresponds to $L_{\rm min} = 4 \times 10^{-5}~L^*$. This is in good agreement with the fiducial value of $L_{\rm min} = 10^{-4}~L^*$ assumed in this work, when taking the significant scatter of the halo mass relations into account.

	The halo properties of DLA galaxies can furthermore be studied by analysing the cross-correlation of DLAs and \lya\ forest absorbers \citep{FontRibera2012, Perez-Rafols2018a}. Based on SDSS DR12, \citet{Perez-Rafols2018a} find that the observed bias is consistent with a minimum halo mass of $M_{\rm min} \sim 10^{9}$~M$_{\odot}$ if the DLA cross-section scales with halo mass as a power-law with index $a\approx1.05$. This might be slightly at odds with the lower range of the power-law index inferred by \citet{Bird2014} although not ruled out. It is on the other hand consistent with the results of our model within the rather large uncertainty on $t$.

	\subsection{Metallicity gradients}
	The best-fit value of the metallicity gradient for $L^*$ galaxies, $\gamma^* = -0.019 \pm 0.008$~dex~kpc$^{-1}$, is in good agreement with the previous measurements of the average metallicity gradient for DLAs of $\gamma=-0.022\pm0.004$ \citep{Christensen2014}.
	This agreement is consistent with the fact that most of the high-redshift galaxies analysed by \citet{Christensen2014} are fairly bright and have luminosities around $L^*$. Similar estimates are reported by \citet{Peroux2012}, although the scatter among individual measurements is significant.
	When studying emission-selected galaxies at $z\gtrsim1$, a large range in metallicity gradients has been observed in galaxy discs \citep[e.g.,][]{Swinbank2012, Stott2014, Curti2020}. \citet{Curti2020} report flat or negative gradients for the majority of galaxies, and only in a few cases do they observe inverted metallicity gradients (i.e., more metal-rich at larger radii). The authors find a slight trend of steeper, negative metallicity gradients for more massive galaxies, contrary to the relation found in our model, where the metallicity gradient {\it flattens} for more luminous (and thus massive) galaxies; $\gamma \propto L^{-0.5}$.
	This disagreement might however be a result of sample selection effects together with the very different ways by which metallicity gradients are measured in absorption and emission as well as the physical scales they probe.
	
	The study by \citet{Curti2020} has very few galaxies at $z>2$, and all of these are highly star-forming (SFR~$\sim50$~M~yr$^{-1}$) and rather massive ($M_{\star} \sim 10^{10}$~M$_{\odot}$). 
	DLA hosts in our model probe faint galaxies with low star-formation activity corresponding to stellar masses in the range of $\lesssim 10^9$~M$_{\odot}$.
	It is thus plausible that the differences derived in the luminosity (or mass) dependence is due to the very different sample characteristics in terms of stellar mass.
	We furthermore note that while luminosity and mass are correlated, it is not a one-to-one correspondence due to variations in star-formation histories and dust attenuation.
	
	Beyond sample selection effects, the emission samples further present a mix of various emission line diagnostics which have complicated systematic effects \citep[e.g.,][]{Kewley2008}. In studies combining emission and absorption, similar systematic effects come into play \citep[see][]{Peroux2012, Rahmani2016}. Lastly, the observations of high-redshift galaxies in emission only probe the inner few kpc of the brightest galaxies. In contrast, our analysis takes into account the average metallicity gradient of the whole DLA galaxy population out to large distances. We thus conclude that the metallicity gradient for DLA galaxies included in our model is not directly comparable to the observed emission-line-derived metallicity gradients, and any differences might therefore be ascribed to differences in sample selection and methodology. Further investigation is needed to analyse these systematics in detail; However, this is beyond the scope of this work.
	
	Both studies by \citet{Curti2020} and \citet{Stott2014} report a correlation between the metallicity gradient and the specific star formation rate, indicating that more vigorously star-forming galaxies have flatter gradients (or even inverted gradients). Isolating more regular galaxies on the so-called `star formation main sequence', \citet{Stott2014} find an average metallicity gradient of $\langle \gamma \rangle = -0.020\pm0.004$ in surprising agreement with our results. This might indicate that the bulk of DLAs does not probe strongly star-bursting galaxies, consistent with the low star formation activity observed in direct detections of high-redshift DLAs \citep{Krogager2017, Rhodin2018}.

	\subsection{Completeness of observations}
	
	As eluded to above, one complication in our modelling is the complex selection effects of the observational data. The \NHI\ data by \citet{Noterdaeme2012c} are fairly complete and homogeneous based on BOSS data release 9. These data are limited by a signal-to-noise criterion in order to have enough signal in the \lya-forest to detect the DLAs. This translates to an effective magnitude limit in the $g$ band. On the other hand, the $Z$ measurements are very heterogeneous as they are obtained by different follow-up campaigns which are often preselected on different and poorly quantified criteria. However, in order to obtain high-resolution data with current 8--10 m class telescopes, the background quasars are required to be brighter than $r \lesssim 20$. Both of these effective magnitude limits on the \NHI\ and $Z$ samples are well-reproduced by the selection probability as implemented in this work. Yet, we caution that the value of \Avcrit\ might be different for the \NHI\ and $Z$ samples.
	The only way to properly overcome these selection effects is by obtaining large and homogeneous samples of DLAs with measurements of both \NHI\ and $Z$.

\section{Summary}
\label{summary}

	In this work, we have presented an extension of the model by \citet{Fynbo2008} with the aim of reproducing the joint distribution of \NHI\ and $Z_{\rm abs}$ for DLAs at redshifts $z = 2-3$. The model assumes that the galaxies giving rise to DLAs are drawn from the population of star-forming galaxies following the UV luminosity function. The effective DLA cross-section, $\sigmaHI$, around each galaxy is assumed to scale with the luminosity of the galaxy: $\sigmaHI = \pi R^2_{\textsc{dla}}$, where $R_{\textsc{dla}}$ scales with luminosity as $R_{\textsc{dla}} \propto L^t$. Furthermore, the galaxies are assumed to follow a metallicity--luminosity relation and exhibit an average radial metallicity gradient.
	
	We have included a simple prescription for the radial column density profile of \HI. We have found that a log-normal scatter around this average radial \HI\ profile is needed in order to match the high \NHI\ tail of the distribution. We furthermore include a selection bias due to dust obscuration in optically selected quasar samples as quantified by \citet{Krogager2019}. The model contains 8 free parameters (and 4 parameters which are kept fixed at their assumed fiducial values) summarized in Table~\ref{tab:pars}. In order to constrain these parameters, we use an MCMC sampler to obtain the posterior probabilities.

	The best-fit model provides a good fit to the data as seen in Fig.~\ref{fig:NHI_Z} and we find that the modelled distribution of impact parameters agrees well with observations even though these were not used to constrain the model. This agreement is highlighted when considering only model galaxies that would be bright enough in the UV to be detected with current facilities (see Fig.~\ref{fig:bNHI_only}).
	
	By converting the UV luminosities in our analysis to halo masses \citep[following][]{Mason2015}, we compare the distribution of DLA galaxy luminosities in our model to the halo mass distribution seen in numerical simulations \citep{Pontzen2008, Bird2014}. We find that the model distribution is consistent with numerical simulations in terms of cross-section and its scaling with luminosity (halo mass) as well as the lower luminosity (mass) limit for DLA cross-section of $\sim10^{-4}~L^*$ ($\sim10^9$~M$_{\odot}$).
	
	Lastly, we find strong evidence for a negative radial metallicity gradient which scales inversely with luminosity, i.e., more luminous galaxies have flatter gradients. While this trend with luminosity is somewhat at odds with the mass dependence seen in observations \citep[e.g.,][]{Curti2020}, the average value of the gradient for $L^*$ galaxies, $\gamma^* = -0.019 \pm 0.008$~dex~kpc$^{-1}$, is in agreement with other works \citep{Christensen2014, Stott2014}.

\section*{acknowledgements}
	We thank the anonymous referee for the very constructive report.
	We would like to thank Matt Lehnert for fruitful discussions.
	The research leading to these results has received funding from the French
	{\sl Agence Nationale de la Recherche} under grant no ANR-17-CE31-0011-01
	(project ``HIH2'' -- PI Noterdaeme).

\bibliographystyle{mnras}

\end{document}